\documentclass[iop]{emulateapj}

\usepackage{color}
\usepackage{graphicx}
\usepackage{amsmath}

\shorttitle{Testing Lorentz Invariance Violation with gamma-ray bursts}
\shortauthors{Pan et al.}

\newcommand{\beq}{\begin{equation}}
\newcommand{\eeq}{\end{equation}}

\newcommand{\lk}{\left}
\newcommand{\rk}{\right}
\def\ba{\begin{eqnarray}}
\def\ea{\end{eqnarray}}

\begin{document}
\title{ Model-independent constraints on Lorentz invariance violation: implication from updated Gamma-ray
burst observations}
\author{Yu Pan$^{1}$, Jingzhao Qi$^{2}$, Shuo Cao$^{3\ast}$, Tonghua Liu$^{3}$, Yuting Liu$^{3}$, Shuaibo Geng$^{3}$, Yujie Lian$^{3}$, and Zong-Hong Zhu$^{3}$}

\altaffiltext{1}{College of Science, Chongqing University of Posts
and Telecommunications, Chongqing 400065, China;}
\altaffiltext{2}{Department of Physics, College of Sciences,
Northeastern University, Shenyang 110004, China;}
\altaffiltext{3}{Department of Astronomy, Beijing Normal University,
100875, Beijing, China; \emph{caoshuo@bnu.edu.cn}}

%

\begin{abstract}

Astrophysical observations provide a unique opportunity to test
possible signatures of Lorentz Invariance Violation (LIV), due to
the high energies and long distances involved. In quantum theory of
gravity, one may expect the modification of the dispersion relation
between energy and momentum for photons, which can be probed with
the time-lag (the arrival time delay between light curves in
different energy bands) of Gamma-ray bursts (GRBs). In this paper,
by using the detailed time-delay measurements of GRB 160625B at
different energy bands, as well as 23 time-delay GRBs covering the
redshifts range of $z=0.168-2.5$ (which were measured at different
energy channels from the light curves), we propose an improved
model-independent method (based on the newly-compiled sample of
$H(z)$ measurements) to probe the energy-dependent velocity due to
the modified dispersion relation for photons. In the framework of a
more complex and reasonable theoretical expression to describe the
time delays, our results imply that the intrinsic time lags can be
better described with more GRBs time delay data. More
importantly, through direct fitting of the time-delay measurements
of a sample of GRBs, our limit on the LIV energy scale is comparable
to that with unknown constant for the intrinsic time lag, much lower
than the Planck energy scale in both linear LIV and quadratic LIV
cases.

\end{abstract}


\section{Introduction}

As one of the significant pillars of special/general relativity and
particle physics, Lorentz invariance, which plays a very important
role in modern physics, has been confirmed in all observations
devoted to its testing to date (especially in solar system and
colliders). However, during the last two decades a great attention
has been paid to many quantum gravity (QG) theories with possible
Lorentz invariance violation (LIV), in which the LIV will happen
above the Planck energy scale ($E_{QG}\approx E_{Pl}=\sqrt{\hbar
c^5/G} \simeq 1.22\times 10^{19} \rm{GeV}$ due to the quantization
of space-time \citep{Mattingly05,Amelino-Camelia13}). Any possible
violation of Lorentz invariance would have far-reaching consequences
for our understanding of the Nature, i.e., the pillars of modern
physics will be shocked and new physics is needed \citep{Cao18a}.
Therefore, the pursuit of testing the possible Lorentz invariance
violation (LIV) at at much higher precision has continued in recent
decades, concerning various astrophysical and cosmological
observations.

Formulating and quantitatively interpreting the test of LIV is
another question: an interesting proposal, in this respect, has been
formulated in the frameworks of many quantum gravity (QG): at small
spatial scales, a foamy structure of space-time predicted by QG
theory will interact only with the high energy photons
\citep{Amelino-Camelia98,Amelino-Camelia01}. In this case, the speed
of light is varying at different energy range in vacuum, i.e., the
high and low energy photons will not reach us at the same time. More
specifically, the deformed velocity of light usually takes the form
$v=c(1-s_{\pm} E/E_{QG})$ where $E_{QG}$ is effective QG energy
scale, $s_{\pm}$ is a dimensionless parameter depending on the the
particular QG model, and $c$ is the limiting speed of light on low
energy scales, respectively. On low energy scales, $E\ll E_{QG}$,
the effect of LIV will be more obvious and the high energy photons
propagate slower than low energy photons. Although the QG effect is
expected to be very weak (since $E_{QG}$ is typically close to the
Planck energy scale), some effects of LIV are expected to increase
with energy and over very long distances due to cumulative processes
in photon propagation. Therefore, astrophysical searches provide
sensitive probes of LIV and its potential signatures, such as the
energy-dependent time delay and many other phenomena. Gamma-ray
bursts (GRBs), the most luminous astrophysical events observable
proposed as distance indicators at high redshift
\citep{Pan13,Cao14}, creates such opportunity to test the Lorentz
invariance violation (LIV), through the precise measurement of
unprecedented very-high-energy photons. Compared with commonly-used
supernova Ia, the advantages of GRB lies in its high energy photons
in the gamma-ray band (from KeV to GeV), short spectral lags, and
the propagation distance at cosmological scales. Such a natural
laboratory provides a possibility of testing LIV, through the
well-measured time delays between light curves in different energy
bands caused by the LIV.

Recently, some advances have been made concerning the limits on LIV
using different samples of GRBs \citep{Ellis03,
Ellis06,Rodriguez06,Jacob07,Wei17a,Pan15,Wei17c,ZhangS15,Zhang18}.
The original idea of such studies of possible LIV constraints can be
traced back to the papers of \citet{Ellis03}, which developed a
method to analyze samples of GRBs with different redshifts and
energy bands, by extracting time-dependent features from the GRB
signals. It is worth noting that the intrinsic time delay and the
linear term denotes LIV effect should be taken into account when one
calculates the observed time delay $\Delta t$ (see Section 2 for
details). Then, \citet{Ellis08,Jacob08} studied the possibility to
test energy-dependent time delays through GRB measurements, which
would result in strong sensitivity limits to LIV in the photon
sector. Although no strong evidence of the LIV is currently
supported by this astrophysical probe, one important issue should be
reminded: the calculated time delay is strongly dependent on the
cosmic expansion history (characterized by the Hubble parameter
$H(z)$) and thus the pre-assumed cosmological model (the standard
$\Lambda$CDM model, etc) in all of the relevant works. For instance,
it was found in \citet{Ellis06} that there is no strong evidence of
LIV and the effective QG energy scale can be determined at $E_{QG}
\geq 1.4 \times 10^{16}$ GeV, in the framework of the concordance
$\Lambda$CDM model with all of the model parameters taken from the
typical results from WMAP observations. A weak evidence of LIV was
also noticed and discussed in \citet{Biesiada09}, which studied the
LIV by applying the quintessence and Chaplygin Gas model to the
observational GRB time delays. However, it should be noted that in
the above analysis all the values of the corresponding model
parameters have been fixed and the degeneracies among cosmological
parameters were neglected. Further progress in this direction has
recently been achieved by \citet{Pan15,Wei17a} in two recent paper,
which respectively constrain different cosmological (or
cosmographical) parameters together with the LIV parameters by using
the observational data. While comparing the results from their
works, no apparent evidence of LIV and weak hints for LIV are
respectively reported. Up to now, there existed several explanations
of this weak hints of possible LIV. First of all, it may be just a
statistical result produced by the limited amount of observational
data available. In order to draw firm and robust conclusions, one
will need to minimize statistical uncertainties by increasing the
depth and quality of observational data sets. Secondly, to our best
knowledge, one or more particular cosmological models have been
assumed in (almost) all of the relevant works in the literature,
which makes the results on LIV in those works model-dependent and
hence not so robust in fact. Therefore, reconstruction of the cosmic
expansion history may strongly influence the estimated values of the
LIV parameters. Thirdly, one cannot ignore the fact that the weak
hints of possible LIV may be brought by some caveats in the LIV
parametrization.

In this context, it is clear that collection of more complete
observational data concerning time-delay measurements does play a
crucial role. The purpose of our paper is to show how the
combination of the most recent and significantly improved time-delay
measurements of GRB 160625B at different energy bands, as well as 34
time-delay GRBs covering the redshifts range of $z=0.168-4.3$ (which
were measured at different energy channels from the light curves)
can be used to probe possible signal of LIV and set limits on the
value of $E_{QG}$ \citep{Ellis06}. More importantly, compared with
the previous works using the luminosity distances from type Ia
supernova, we will use instead, cosmological distances covering the
GRB redshift range derived in a cosmological-model-independent way
from Hubble parameter measurements using Gaussian processes (GP),
based on the newly-compiled sample of $H(z)$ measurements
\citep{Wei17b}. In order to discuss the LIV in a general framework,
a more complex and reasonable theoretical expression will be
considered in our analysis. We expect that the newest measurements
of GRBs combined with non-parametric distance reconstruction from
the most recent $H(z)$ data will shed much more light on the
possible hint for LIV at higher redshifts. This paper is organized
as follows. In Section~\ref{sec:LIV}, we briefly describe the
methodology of the intrinsic time lag and the time delay induced by
LIV. Then, in Section~\ref{sec:GP} we introduce the time delay data
from GRBs, and the observational data of the Hubble parameters used
in our analysis. The results and corresponding discussion
are presented in Section 4. Finally, we summarize our conclusions in
Section 5.



\section{The Lorentz Invariance Violation } \label{sec:LIV}

As one of the successful predictions of general relativity in the
past decades, the energy dependent velocity of light is
$E^2=p^2c^2$, by assuming constant speed of light and the validity
of Lorentz invariance. However, on low energy scales the
introduction of a Lorentz violating term in the standard model can
induce modifications to the particle dispersion relation
\begin{equation}
E^2=p^2c^2\left[1-s_{\pm}\left(\frac{E}{E_{QG}}\right)^n\right],
\end{equation}
where $s_{\pm}=\pm 1$ ($s_{\pm}=+1$ or $s_{\pm}=-1$ stands for a
decrease or an increase in photon velocity with an increasing photon
energy), while the $n$ parameter represents the leading order
(linear term or quadratic term) of the correction from the
underlying theory ($n=1$ corresponds to the Double Special
Relativity \citep{Magueijo02,Amelino-Camelia02,Amelino-Camelia10}
and $n=2$ corresponds to Extra-Dimensional Theories
\citep{Sefiedgar11} or Harava-Lifshitz Gravity
\citep{Horava09a,Horava09b,Vacaru10,Blas11}). Now the
energy-dependent speed of photon can be written as
\begin{eqnarray}
v=\frac{\partial E}{\partial p}=c\left[1-s_\pm\frac{n+1}{2}\left(\frac{E}{E_{QG}}\right)^n \right]
\end{eqnarray}
Note that we only consider the case of $s_\pm=+1$ in the present
work, since the high-energy photons travel slower than their
low-energy counterparts. On the other hand, the linear term ($n=1$)
obviously dominate the dispersion relation for $E \ll E_{QG}$. In
order to have a better extension and discussion, in this analysis we
also consider the second case with quadratic term ($n=2$), which is
different from most of the relevant works in the literature.

Now over long distances due to cumulative processes in photon
propagation, one can get the LIV-induced time delay between photons
with high energy and low energy
\citep{Jacob07,Jacob08,Biesiada07,Biesiada09}
\begin{equation}
\Delta t_{\rm{LIV}}=
\frac{1+n}{2H_0}\frac{E^n-E^n_0}{E^n_{QG,n}}\int_0^z\frac{(1+z')^ndz'}{h(z')}
\label{time delay}
\end{equation}
where $H_0$ is the present value of the Hubble function, $h(z) =
H(z)/H_0$ is a dimensionless expansion rate dependent on redshift
$z$. From observational point of view, for a cosmic transient source
(e.g. GRB), the observed time delay between two different energy
bands should include five terms
\begin{equation}
\Delta t=\Delta t_{\rm{LIV}}+\Delta t_{\rm{int}}+\Delta
t_{\rm{spe}}+\Delta t_{\rm{DM}}+\Delta t_{\rm{gra}}
\end{equation}
where $\Delta t_{\rm{spe}}$ is related to the potential time delay
due to special relativistic effects if photons have a non-zero rest
mass \citep{Gao15}, while $\Delta t_{\rm{DM}}$ and $\Delta
t_{\rm{gra}}$ respectively denote the time delay contribution from
the dispersion by the line-of-sight free electron content and the
gravitational potential along the propagation path of photons if the
Einstein¡¯s equivalence principle (EEP) is violated \citep{Wei15}.
However, the effect of the three terms is negligible for GRB
photons, following the recent analysis of \citet{Wei15,Gao15}. The
first term, $\Delta t_{\rm{LIV}}$ represents the LIV-induced time
delay, while the second term $\Delta t_{\rm{int}}$ quantifies the
intrinsic time delay describing that photons with high and low
energies do not leave the source simultaneously. In order to account
for the unknown intrinsic time lags, the unknown intrinsic time
delays of GRBs are usually specified by including a parameter $b$ in
the rest-frame of the source, which can be written as $\Delta
t_{\rm{int}}=b(1+z)$ when the cosmic expansion is taken into account
\citep{Ellis06,Pan15}. More recently, it was proposed in the recent
analysis that the assumption that the intrinsic time lag between the
lowest energy band and any other high energy bands increases with
the energy $E$ is more reasonable \citep{Wei17c}. In this paper we
will adopt this more reasonable formulation of the intrinsic
energy-dependent time lag, in the form of an approximate power-law
function:
\begin{equation}
\Delta
t_{\rm{int}}(E)=\tau\lk[\lk(\frac{E}{\rm{keV}}\rk)^\alpha-\left(\frac{E_0}{\rm{keV}}\right)^\alpha\right],
\end{equation}
where $\tau$ and $\alpha$ are two positive free parameters, and
$E_0$ is the low energy band. Therefore, the theoretical time delay
between different energy bands of GRBs can be expressed as
\begin{eqnarray} \nonumber
\Delta t_{\rm{th}}&=&\Delta t_{\rm{LIV}}+\Delta t_{\rm{int}}(1+z) \\
&=& \frac{1+n}{2H_0}\frac{E^n-E^n_0}{E^n_{QG,n}}\int_0^z\frac{(1+z')^ndz'}{h(z')} \nonumber\\
&&+\tau\lk[\lk(\frac{E}{\rm{keV}}\rk)^\alpha-\left(\frac{E_0}{\rm{keV}}\right)^\alpha\right]\lk(1+z\rk).
\end{eqnarray}
Note that there are three parameters in this model \{$E_{QG}$,
$\tau$, $\alpha$\}. For convenience, we define a parameter $K(z)$ as
\citep{Pan15}
\begin{equation}
\label{Kzeq}
K(z)=\int_0^z\frac{(1+z')^ndz'}{h(z')},
\end{equation}
which is related to the measurements of cosmic distances. As is
mentioned above, almost all of the previous test of LIV used the
cosmological models to calculate the distance-like parameter $K(z)$,
based on the General Relativity frame without LIV. Therefore, a more
reasonable approach will be applied to construct $K(z)$, which is
model-independent, rather than assuming a particular cosmo-logical
model as in the literature.

\begin{figure}
\includegraphics[width=1.0\hsize]{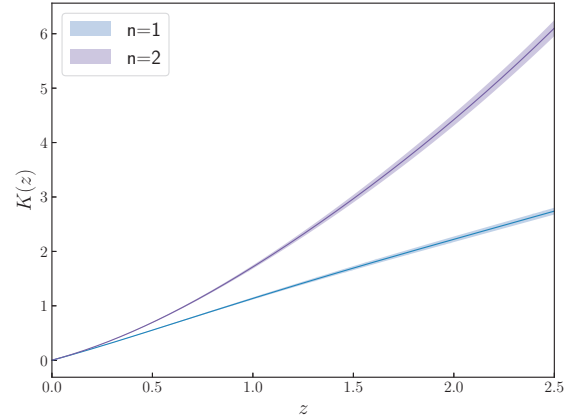}
\caption{ The distribution of reconstructed $K(z)$ with redshifts,
in the two cases of $n=1$ and $n=2$. \label{Kz}}
\end{figure}

\begin{table}
\begin{center} \caption{The 1$\sigma$ constraints on $E_{QG}$, $\tau$ and
$\alpha$ for different LIV models ($n=1, n=2$).}
\begin{tabular}{|c|c|c|} \hline
Model parameter & $n=1$ & $n=2$ \\
\hline
$\log{E_{QG}}(GeV) $& $14.523_{-0.025}^{+0.022}$ & $8.79_{-0.0097}^{+0.0097}$ \\
\hline
$\tau$& $0.00067_{-0.00005}^{+0.00005}$ & $0.0000018_{-0.00000009}^{+0.00000009}$ \\
\hline
$\alpha$& $1.0556_{-0.0046}^{+0.0046}$ & $2.0209_{-0.0034}^{+0.0034}$ \\
\hline
\end{tabular}\label{all}
\end{center}
\end{table}

\section{Observational data} \label{sec:GP}

In this paper, we use the time delay data of GRB 160625B at
different energy bands (whose redshift is $z=1.41$) \citep{Xu16} and
34 time-delay GRBs (with the redshifts spanning from $z=0.168$ to
$z=4.3$), which were derived from time lags between different energy
channels measured from the light curves.

The GRB 160625B was triggered and located twice by the Fermi
Gamma-Ray Burst Monitor (GBM) \citep{Burns16}, with a sharp increase
in the rate of high-energy photons and an onboard trigger on a
bright pulse detected by the fermi Large Area Telescope (LAT). The
gamma-ray light curve of GRB 160625B consists of three dramatically
different isolated sub-bursts, while the spectral time lags are
obtained from the light curves in the 15 - 350 keV energy band with
respect to a total duration of $\sim T_{90}=770s$ \citep{Zhang16b}.
One should note that the second sub-burst of GRB 160625B is very
bright, which made it possible to easily extract its light curves in
different energy bands. The observed time lags measured from the
energy-dependent light curves are listed in Table 1 of
\citet{Wei17c}. Meanwhile, based on the techniques from signal
processing such as wavelet analysis to identify and correlate
genuine features in the intensities observed in different energy
bands \citep{Ellis03}, a time delay data set from 35 GRBs was
compiled by \citet{Ellis06}, with known redshifts from $z=0.168$ to
$z=6.29$. All of the data were shown in Table 1 of \citet{Ellis06}
including the 9 time-delay of GRBs from light curves whose time
resolution is 64 ms and redshifts span from $z=0.835$ to $z=3.9$
observed by BATSE spectral channels, the HETE data with 15 light
curves whose time resolution is 164 ms and redshifts span from
$z=0.168$ to $z=3.372$, and the SWIFT data with 10 light curves
whose time resolution is 64 ms and redshifts span from $z=0.258$ to
$z=4.3$. The spectral time lags are obtained from the light curves
in the 25 - 320 keV energy band. The observed time lags measured
from the energy-dependent light curves are listed in Table 1 of
\citet{Ellis06}.

On the other hand, we also use Gaussian processes to reconstruct the
function $K(z)$ from observational $H(z)$ data directly. Such idea
was first discussed in \citet{Holsclaw10} and then extensively
applied in more recent papers to test the cosmological parameters
\citep{Cao17a,Cao18}, spatial curvature of the Universe
\citep{Cao19,Qi19a}, and the speed of light at higher redshifts
\citep{Cao17b}. In this process, the reconstructed function $f(x)$
at different points $x$ and $\tilde{x}$ are correlated by a
covariance function $k(x,\tilde{x})$ \citep{SeikeL12a}. The commonly
used function is the squared exponential covariance, whose advantage
is infinitely differentiable, which only depends on two hyper
parameters $\ell$ and $\sigma_f$. Both $\ell$ and $\sigma_f$ would
be trained to determine the specific value  by GP code self with the
observational data. Therefore, the GP method does not specify any
form of $f(x)$ and is model-independent. Moreover, we use the
publicly available code called the GaPP (Gaussian Processes in
Python) \footnote{http://www.acgc.uct.ac.za/~seikel/GAPP/index.html}
reconstruct the profile of $H(z)$ function up to the redshifts
$z=2.5$, which has been widely used in various studies
\citep{Yang15,Qi16,Zhang16a}. In this paper, we use the newest
Hubble parameter ($H(z)$) data set, which consist of 31 measurements
from the differential ages of passively evolving galaxies and 10
measurements via the detection of radial BAO features
\citep{Zheng19}. One should note that the limited redshift range of
$H(z)$, i.e., $z\sim2.5$ implies that only a limited number of known
GRBs can be used. This selection leaves us with the detailed
time-delay measurements of GRB 160625B \citep{Wei17c} and 23
time-delay GRBs summarized in \citet{Ellis06}.

\section{Results and discussion}

Using the aforementioned GP, one is able to reconstruct the profile
of $K(z)$ function up to the redshift of $z=2.5$ (assuming flat
universe). The results are shown in Fig.~\ref{Kz}, in the framework
of two cases with $n=1$ and $n=2$. Based on the reconstructed
functions of $K(z)$, we determine the parameters ($E_{QG}$, $\tau$
and $\alpha$) characterizing LIV by minimizing the $\chi^2$
objective function
\begin{equation}
\chi^2_{GRB}=\sum_{i=1}^{N_{GRB}}\left[\frac{\Delta t_{th}(E_{QG},
\tau, \alpha)-\Delta t_{obs}}{\sigma_{\Delta t}}\right]^2,
\end{equation}
where $\Delta t_{th}$ denotes the theoretical time delays of GRB,
and $\Delta {t_{obs}}$ is the observational counterpart with the
corresponding 1$\sigma$ uncertainty ($\sigma_{\Delta t}$). We apply
the Monte Carlo Markov Chain (MCMC) method \citep{Lewis02} with 8
chains and obtain the marginalized 1$\sigma$ and 2$\sigma$
constraints. Performing fits to different scenarios ($n=1$ and
$n=2$) on the GRB sample, we obtain the results displayed in Table
1. The marginalized probability distribution of each parameter and
the marginalized 2-D confidence contours are presented in Figs.~1-2.

\begin{figure}
\includegraphics[width=0.9\hsize]{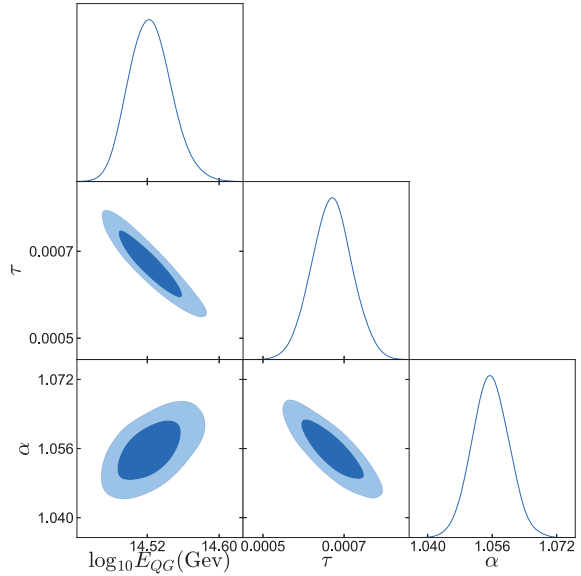}
\caption{The 1D probability distribution of each parameter and the
2D confidence contours for the parameters $E_{QG}$, $\tau$ and
$\alpha$ (the linear LIV case, i.e., $n=1$). \label{livn1}}
\end{figure}

\begin{figure}
\includegraphics[width=0.9\hsize]{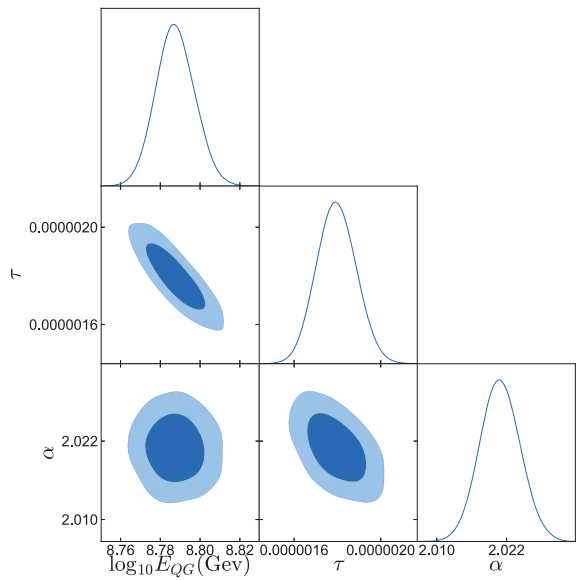}
\caption{The 1D probability distribution of each parameter and the
2D confidence contours for the parameters of $E_{QG}$, $\tau$ and
$\alpha$ (the quadratic LIV case, i.e., $n=2$). \label{livn2}}
\end{figure}

Let's start from the first case by fitting the observed time-delay
data with the linear LIV case (i.e., $n=1$). Performing fits on the
the time-lag (the arrival time delay between light curves in
different energy bands) of GRBs, we obtain the following best-fit
values and corresponding 1$\sigma$ uncertainties (68.3\% confidence
level):
\begin{eqnarray}
&& \rm {log} {E_{QG,1} \rm(GeV)}=14.523_{-0.025}^{+0.022}, \nonumber\\
&& \tau=0.00067_{-0.00005}^{+0.00005}, \nonumber\\
&& \alpha=1.0556_{-0.0046}^{+0.0046}. \nonumber
\end{eqnarray}
The resulting constraints on $E_{QG,1}$, $\tau$ and $\alpha$ are
shown in Fig.~\ref{livn1}. Thanks to the improved
statistical technique and the use of a more complete data set, the
result of our present analysis is significantly stronger and more
robust than that in \citet{Ellis06}.Since Lorentz invariance plays
an important role in modern physics, it is interesting to to
distinguish the possible time delay induced by the LIV effect from
any source-intrinsic time lag in the emission of photons at
different energies.  On the one hand, tighter estimates of the
intrinsic energy-dependent time lag is obtained. We find that the
best-fitted value of the parameter $\tau$ is a small number,
$|\tau|\sim 6\times10^{-4}$. On the other hand, different from the
previous procedure assuming an unknown constant to characterize the
intrinsic time lag for all GRBs, our results show that the intrinsic
lag has a positive dependence on the photon energy. More
specifically, compared with the previous results obtained on the
individual source GRB 160625B \citep{Wei17c}, our full GRB sample
analysis has yielded improved constraints on this meaningful
physical parameter, $\alpha\sim1.05$. It should be noted that in the
framework of LIV, such positive correlation between the lag and the
energy will gradually become an anticorrelation at the high energy
scales, since high energy photons travel slower than low energy
photons in vacuum \citep{Wei17c}. More importantly, more rigorous
quantitative analysis supports the limit of the QG energy scale on
the linear LIV case, $E_{QG,1}\geq0.3\times10^{15}$GeV, four orders
of magnitude below the Planck energy scale. Such analysis is
comparable to the limit found from \citet{Ellis06} with unknown
constant for the intrinsic time lag for the linear LIV case, through
direct fitting of the time-delay measurements of a sample of GRBs.

Next, we consider the quadratic LIV case (i.e., $n=2$) to fit the
observed time-delay data data. Focusing on the second case with the
quadratic term, the best fit is
\begin{eqnarray}
&& \rm {log} {E_{QG,2} \rm(GeV)}=8.79_{-0.0097}^{+0.0097}, \nonumber\\
&& \tau=0.0000018_{-0.00000009}^{+0.00000009}, \nonumber\\
&& \alpha=2.0209_{-0.0034}^{+0.0034}. \nonumber
\end{eqnarray}
Marginalized probability distributions for each parameter and
marginalized 2D 68\% confidence contours are presented in Fig.~2.
The estimation of these LIV parameters is briefly summarized in
Table 1. Comparing constraints based on a different value
for the $n$ term, we see the two parameters quantifying the
intrinsic time lag in different energy bands are in general
agreement with the corresponding quantities for the linear LIV case.
More specifically, the best-fit value of $\tau$ still tends to be
zero ($|\tau|\sim 6\times10^{-6}$), while its positive dependency on
the photon energy ($\alpha\sim 2.02$) is consistent with the
observational data within 1$\sigma$ confidence region. In broad
terms, this reveals that both cases are adequate to represent the
time lag of GRBs. With the best-fit values of $\rm {log} {E_{QG,2}
\rm(GeV)}$ as well as its 1$\sigma$ error bar, the 1$\sigma$
confidence-level lower limit on LIV is
$E_{QG,2}\geq0.6\times10^{9}$GeV and the sensitivity of this
analysis might have been expected to be two orders of magnitude
greater than in \citet{Wei17c}. In this way, our results can exclude
the energy scale of the new LIV physics, $E_{QG,2}$, to greater than
$10^9$GeV. The most conservative limit on the violation of Lorentz
invariance that we find is two orders of magnitude below the current
best limit from the single GeV photon of GRB 090510
\citep{abdo09b,Vasileiou13}. Note that, although some stronger upper
limits on a modification of the photon dispersion relation have been
reported in the literature, any analysis of a single source can only
be regarded as indicative, due to unknown systematic uncertainties
associated with unknown intrinsic spectral properties of any given
GRB \citep{Ellis06}. In order to establish a rigorous limit, one
must focus on the differentiation between intrinsic and propagation
effects in the time-delay measurements, which can be done robustly
only by analyzing a sizeable statistical sample of GRBs. Now it is
worthwhile to make some general comments on the reliability of the
resulting constraints on LIV. On the one hand, the analysis of the
intrinsic time lag performed here is important for studying the
flight time differences from the astronomical sources to test the
LIV effect. Different from the previous model with an unknown
constant for the intrinsic time lag \citep{Ellis06}, we use a more
complex and reasonable theoretical expression to describe the
intrinsic energy-dependent time lags \citep{Wei17c}. Meanwhile,
compared with the previous works using the luminosity distances from
type Ia supernova to quantify LIV effect \citep{Pan15}, we use
instead, cosmological distances derived in a cosmological
model-independent way from $H(z)$ measurements using Gaussian
processes (GP) \citep{Wei17b,Zheng19}. By considering the
contribution of both the intrinsic time lag and the lag by the LIV
effect, our analysis indicates that the intrinsic time lags can be
obviously better described with more GRBs time delay data. More
importantly, it is possible to give robust limits on LIV through
direct fitting of the spectral lag data of a GRB.

There are several sources of systematics that we do not
consider in this paper and remain to be addressed in future
analysis. First of all, we have learned from our analysis that the
reliability of the resulting constraints on LIV strongly depends on
a good knowledge of the intrinsic time lag, i.e., photons with high
and low energies do not leave the GRB simultaneously. Although the
problems associated with the intrinsic time delay can be handled
better with the new formulation of $\Delta t_{int}$, such
astrophysical term is still difficult to predict since it depends
only on good understanding of the physics of source evolution
\citep{Ellis08}. More importantly, the strong correlation between
the parameters describing intrinsic and propagation time-lags have
revealed new systematic issues, which can be clearly seen from Fig.
2-3. Therefore, in our approach the available GRB data can be used
conservatively to set a lower limit on any Lorentz Invariance
Violation (LIV). Further progress in this direction has recently
been achieved by \citet{Ellis06}, which suggested that a supposedly
more reliable subs-ample of GRBs with a spread of different measured
redshifts are advantageous, from the point of view of demonstrating
the absence of any destructive interference between intrinsic and
propagation effects. In this aspect, the reduction of the above
potential systematic bias should turn to better understanding of the
internal dynamics of GRBs or available data extending to much higher
energies \citep{Ellis08}. Secondly, the other source of systematic
uncertainty comes from the reconstruction of the function $K(z)$.
Different from the previous studies assuming a particular
cosmological model (e.g. $\Lambda$CDM) \citep{Wei17c}, we have
applied one particular non-parametric method based on Gaussian
processes to reconstruct the function $K(z)$ from observational
$H(z)$ data. The only way to minimize this unknown systematic
uncertainty is to search for more efficient distance reconstruction
technique or model-independent methodology. This problem has also
been recognized and discussed in many recent works \citep{Zou18},
with a heuristic suggestion that with the so-called cosmography (one
of the powerful model-independent approaches), one can analyze the
evolution of the universe without assuming any underlying
theoretical model. As a final remark, we point out that the
time-lags discussed in this paper are derived from the sharp
features observed in the intensities of radiation with different
energies, identified by various wavelet techniques for different GRB
sub-samples. Therefore, the follow up engaging the observations of
more GRBs with higher temporal resolutions and more high energy
photons may make it less susceptible to such systematic errors.

\section{Conclusions}

Modern ideas in quantum gravity predict the possibility of
Lorenz Invariance Violation (LIV), which reveals itself by energy
dependent modification of standard relativistic dispersion relation.
Following this direction, time of flight delays in photons emitted
by astrophysical sources located at cosmological distances can
become a valuable tool for setting limits on LIV theories. In this
paper we discuss an improved model-independent method to constrain
the energy-dependent velocity due to the modified dispersion
relation, based on the detailed time-delay measurements of GRB
160625B at different energy bands and 23 time-delay GRBs covering
the redshifts range of $z=0.168-2.5$. Two parametric expressions are
considered to describe the LIV-induced time delay between photons
with high energy and low energy (linear and quadratic LIV cases).
Here we summarize our main conclusions in more detail:

\begin{itemize}

\item In most of the relevant works on LIV, the intrinsic time-lag of GRBs is actually oversimplified by
assuming $\Delta t_{int}=b(1+z)$, which means that all GRBs have the
same intrinsic time delay in the source frame. Meanwhile, one or
more particular cosmological models have been assumed in the
literature, which makes the results on LIV model-dependent. In this
paper, we turn to a more complex and reasonable theoretical
expression to describe the energy-dependent intrinsic time-lag,
while the cosmic expansion history in the LIV time delay is
reconstructed from observational $H(z)$ data based on
model-independent Gaussian processes.

\item Our results show that the intrinsic time lags can be better
described with more GRBs time delay data. Instead of assuming an
unknown constant for the intrinsic time lag, we argue that the
intrinsic lag has a positive dependence on the photon energy for
both linear LIV and quadratic LIV cases. More importantly, the
strong correlation between the parameters describing intrinsic and
propagation time-lags have been revealed in our analysis, which
indicates that a more realistic assumption for the intrinsic
time-lag of GRBs is important to robustly constrain the possible
LIV. In the linear LIV and quadratic LIV cases, one can limit the
LIV energy scale at the level of much lower than the Planck energy
scale: $E_{QG,1}\geq0.3\times10^{15}$GeV and
$E_{QG,2}\geq0.6\times10^{9}$GeV. Although some stronger upper
limits on LIV have been reported from the single GeV photon of a
single source, our approach can be used conservatively to set a
lower limit on any LIV.

\item In conclusion, our analysis demonstrates that time-lags in emissions
from GRBs can already now be used to set limit on the Violation of
Lorentz Invariance, in the framework of a more complex and
reasonable formulation to describe the LIV effect. One may say that
the approach initiated in \citet{Ellis03,Ellis08,Jacob08} can be
further developed to analyze a larger sample of GRBs with different
redshifts and energy bands. Fits on the phenomenological formula
obtained in our analysis, if confirmed by future investigation of
GRB time-delay observations, will offer additional constraints for
possible LIV with extragalactic sources.

\item Finally, it is reasonable to expect that more GRBs with higher temporal
resolutions and more high energy photons (which induce large time
delays) can be used to test the possible LIV. In order to establish
a rigorous limit, we also pin hope on a significantly larger sample
of LIV probes at much higher redshifts, including the photon
time-delay measurements from objects like Active Galactic
Nuclei(AGN) \citep{Albert08} and gravitational waves
\citep{Passos16}. With such complementary probes, combined with
Hubble parameter measurements obtained covering the redshifts range
of $0.1\leq{z}\leq{5.0}$ in the near future \citep{Yu16,Weinberg13},
we can further investigate constraints on the LIV effect and
eventually probe the deep physics behind LIV, i.e., the quantum
gravity theories (string theory, loop quantum gravity, and doubly
special relativity) and field theory frameworks for LIV (the
so-called Standard-Model Extension)
\citep{Kostelecky08,Kostelecky11,Kislat15}.

\end{itemize}

\section*{Acknowledgments}

This work was supported by National Key R\&D Program of China No.
2017YFA0402600, the National Natural Science Foundation of China
under Grants Nos. 11690023, 11373014, and 11633001, the Strategic
Priority Research Program of the Chinese Academy of Sciences, Grant
No. XDB23000000, the Interdiscipline Research Funds of Beijing
Normal University, and the Opening Project of Key Laboratory of
Computational Astrophysics, National Astronomical Observatories,
Chinese Academy of Sciences. J.-Z. Qi was supported by China
Postdoctoral Science Foundation under Grant No. 2017M620661. M.
Biesiada was supported by Foreign Talent Introducing Project and
Special Fund Support of Foreign Knowledge Introducing Project in
China. He is also grateful for support from Polish Ministry of
Science and Higher Education through the grant DIR/WK/2018/12. Y.
Pan was supported by the Scientific and Technological Research
Program of Chongqing Municipal Education Commission (Grant no.
KJ1500414); and Chongqing Municipal Science and Technology
Commission Fund (cstc2015jcyjA00044, and cstc2018jcyjAX0192).


\begin{thebibliography}{0}%
\makeatletter
\providecommand \@ifxundefined [1]{%
 \@ifx{#1\undefined}
}%
\providecommand \@ifnum [1]{%
 \ifnum #1\expandafter \@firstoftwo
 \else \expandafter \@secondoftwo
 \fi
}%
\providecommand \@ifx [1]{%
 \ifx #1\expandafter \@firstoftwo
 \else \expandafter \@secondoftwo
 \fi
}%
\providecommand \natexlab [1]{#1}%
\providecommand \enquote  [1]{``#1''}%
\providecommand \bibnamefont  [1]{#1}%
\providecommand \bibfnamefont [1]{#1}%
\providecommand \citenamefont [1]{#1}%
\providecommand \href@noop [0]{\@secondoftwo}%
\providecommand \href [0]{\begingroup \@sanitize@url \@href}%
\providecommand \@href[1]{\@@startlink{#1}\@@href}%
\providecommand \@@href[1]{\endgroup#1\@@endlink}%
\providecommand \@sanitize@url [0]{\catcode `\\12\catcode `\$12\catcode
  `\&12\catcode `\#12\catcode `\^12\catcode `\_12\catcode `\%12\relax}%
\providecommand \@@startlink[1]{}%
\providecommand \@@endlink[0]{}%
\providecommand \url  [0]{\begingroup\@sanitize@url \@url }%
\providecommand \@url [1]{\endgroup\@href {#1}{\urlprefix }}%
\providecommand \urlprefix  [0]{URL }%
\providecommand \Eprint [0]{\href }%
\providecommand \doibase [0]{http://dx.doi.org/}%
\providecommand \selectlanguage [0]{\@gobble}%
\providecommand \bibinfo  [0]{\@secondoftwo}%
\providecommand \bibfield  [0]{\@secondoftwo}%
\providecommand \translation [1]{[#1]}%
\providecommand \BibitemOpen [0]{}%
\providecommand \bibitemStop [0]{}%
\providecommand \bibitemNoStop [0]{.\EOS\space}%
\providecommand \EOS [0]{\spacefactor3000\relax}%
\providecommand \BibitemShut  [1]{\csname bibitem#1\endcsname}%
\let\auto@bib@innerbib\@empty
\end{thebibliography}%


\begin{thebibliography}{99}
\bibitem[Abdo et al.(2009a)]{abdo09a} Abdo, A. A., Ackermann, M., Ajello, M., et al. 2009, Nature, 462, 331
\bibitem[Abdo et al.(2009b)]{abdo09b} Abdo, A. A., Ackermann, M., Arimoto, M., et al. 2009, Science, 323, 1688
\bibitem[Ade et al.(2014)]{Planck1} Abdo, A. A., Ackermann, M., Arimoto, M., et al. 2014, A\&A, 571, A16
\bibitem[Albert et al.(2008)]{Albert08} Albert, J., Ellis, J., Mavromatos, N. E., et al. 2008, PLB, 668, 253
\bibitem[Amanullah et al.(2010)]{Amanullah10} Amanullah, R., Lidman, C., Rubin, D., et al. 2010, ApJ, 716, 712
\bibitem[Amelino-Camelia.(2010))]{Amelino-Camelia10} Amelino-Camelia, G., World Scientific Publishing Co. Pte. Ltd., 2010. ISBN:9789814287333, 123-170
\bibitem[Amelino-Camelia et al.(1998)]{Amelino-Camelia98} Amelino-Camelia, G., Ellis, J., Mavromatos, N. E., Nanopoulos, D. V., Subir S. 1998, Nature, 393, 763
\bibitem[Amelino-Camelia(2001)]{Amelino-Camelia01a} Amelino-Camelia, G. 2001, PLB, 510, 255
\bibitem[Amelino-Camelia \& Piran(2001)]{Amelino-Camelia01} Amelino-Camelia, G., Piran, T. 2001, PRD, 64, 036005
\bibitem[Amelino-Camelia(2002)]{Amelino-Camelia02} Amelino-Camelia, G. 2002, IJMPD, 11, 35
\bibitem[Amelino-Camelia et al.(2009)]{Amelino-Camelia09} Amelino-Camelia, G., Laemmerzahl, G., Mercati, F., Guglielmo M. T. 2009, PRL, 103, 171302
\bibitem[Amelino-Camelia(2013)]{Amelino-Camelia13} Amelino-Camelia, G. 2013, Living Rev. Rel, 16, 5
\bibitem[Armendariz-Picon et al.(2001)]{Armendariz-Picon01} Armendariz-Picon, C., Mukhanov, V., Steinhardt, P. J. 2001, PRD, 63, 103510
\bibitem[Astier et al.(2006)]{Astier06} Astier, P., Guy, J., Regnault, N., et al. 2006, A\&A, 447, 31
\bibitem[Baukh et al.(2007)]{Baukh07} Baukh, V., Zhuk, A., Kahniashvili, T. 2007, PRD, 76, 027502
\bibitem[Blas \& Sanctuar(2011)]{Blas11} Blas, D., Sanctuary, H. 2011, PRD, 84, 064004.
\bibitem[Beutler et al.(2011)]{Beutler11} Beutler, F., Blake, C., Colless, M., et al. 2011,MNRAS, 416, 3017
\bibitem[Biesiada \& Pi{\'o}rkowska(2007)]{Biesiada07} Biesiada, M., Pi{\'o}rkowska, A. 2007, JCAP, 0705, 011
\bibitem[Biesiada \& Pi{\'o}rkowska(2009)]{Biesiada09} Biesiada, M, Pi{\'o}rkowska, A. 2009, CQG, 26, 125007
\bibitem[Biller et al.(1999)]{Biller99} Biller, S. D., Breslin, A. C., Carson, M., et al. 1999, PRL, 83, 2108
\bibitem[Blake et al.(2011)]{Blake11b} Chris, B., Tamara, D., Gregory, B. P., David, P., Sarah, B. 2011, MNRAS, 418, 1707
\bibitem[Borriello et al.(2013)]{Borriello13} Borriello, E., Chakraborty, S., Mirizzi, A., Serpico, P. D. 2013, PRD, 87, 116009
\bibitem[Burns(2016)] {Burns16} Burns, E. 2016, GRB Coordinates Network, 19581
\bibitem[Caldwell et al.(1998)]{Caldwell98} Caldwell, R. R., Dave, R.,Steinhardt, P. J. 1998, PRL, 80, 1582
\bibitem[Caldwell(2002)]{Caldwell02} Caldwell, R. R. 2002, PLB, 545, 23
\bibitem[Caldwell et al.(2003)]{Caldwell03} Caldwell, R. R., Kamionkowski, M., Weinberg, N. N. 2003, PRL,, 91, 071301
\bibitem[Cao \& Zhu(2011)]{Cao11} Cao, S., Zhu, Z. H. 2011, Science in China G: Physics and Astronomy, 54, 2260
\bibitem[Cao et al.(2012a)]{Cao12a} Cao, S., Pan, Y., Biesiada, M., Godlowski, W.,Zhu, Z. H. 2012a, JCAP, 1203, 016
\bibitem[Cao et al.(2012b)]{Cao12b} Cao, S., Covone, G., Zhu, Z.-H. 2012b, ApJ, 755, 31
\bibitem[Cao \& Zhu(2014)]{Cao14} Cao, S., Zhu, Z.-H. 2014, PRD, 90, 083006
\bibitem[Cao et al.(2017a)] {Cao17a} Cao, S., Biesiada, M., Jackson, J., et al. 2017, JCAP, 02, 012
\bibitem[Cao et al.(2017b)] {Cao17b} Cao, S., Zheng, X. G.,Biesiada, M., et al. 2017, A\&A, 606, A15
\bibitem[Cao et al.(2018a)] {Cao18a} Cao, S., Qi, J. Z., Biesiada, M., et al. 2018a, ApJ, 867, 50
\bibitem[Cao et al.(2018b)] {Cao18} Cao, S., Zheng, X. G., Biesiada, M., Qi, J. Z., Chen, Y. 2018, EPJC, 78, 749
\bibitem[Cao et al.(2019)] {Cao19} Cao, S., Qi, J. Z., Biesiada, M., et al. 2019, PDU, 24, 100274
\bibitem[Chevallier \& Polarski(2001)]{Chevallier01} Chevallier, M., Polarski, D. 2001, IJMPD, 10, 213
\bibitem[Chiba(2002)]{Chiba02} Chiba, T. 2002, PRD, 66, 063514
\bibitem[Eisenstein \& Hu(1998)]{Eisenstein98} Eisenstein, D. J., Hu, W. 1998, ApJ, 496, 605
\bibitem[Eisenstein et al.(2005)]{Eisenstein05} Eisenstein, D. J., Zehavi, I., Hogg. D. W, et al. 2005, ApJ,, 633, 560
\bibitem[Ellis et al.(2003)]{Ellis03} Ellis, J., Mavromatos, N. E., Nanopoulos, D. V., Sakharov, A. S. 2003, A\&A, 402, 409
\bibitem[Ellis et al.(2006)]{Ellis06} Ellis, J., Mavromatos, N. E., Nanopoulos,D.V., Sakharov, A. S., Sarkisyan, E. K. G. 2006, Astroparticle Physics, 25, 402
\bibitem[Ellis et al.(2008)]{Ellis08} Ellis, J., Mavromatos, N. E., Nanopoulos, D. V., Sakharov, A. S. 2008, A\&A, 402, 409
\bibitem[Feng et al.(2005)]{Feng05} Feng, B., Wang, X. L., Zhang, X. M. 2005a, PLB, 607, 35
\bibitem[Feng et al.(2006)]{Feng06} Feng, B., Li, M. Z., Piao, Y. S., Zhang, X. M. 2006, PLB, 634, 101
\bibitem[Gao \& Gong (2014)]{Gao14} Gao, Q., Gong, Y. 2014, CQG, 31, 105007
\bibitem[Gao et al.(2015)]{Gao15} Gao, H., Wu, X. F., M{\'e}sz{\'a},P. 2015, ApJ, 810, 121
\bibitem[Gong et al.(2013)]{Gong12} Gong, Y. G., Gao, Q., Zhu, Z.-H. 2013, MNRAS, 430, 3142
\bibitem[Gong \& Gao(2014)]{Gong13}Gong, Y. G., Gao, Q. 2014, EPJC, 74, 2729
\bibitem[Guo et al.(2005)]{Guo05} Guo, Z. K., Piao, Y. S., Zhang, X. M., Zhang, Y. Z. 2005, PLB, 608, 177
\bibitem[Guo et al.(2012)]{Guo12} Guo, Z. K., Huang, Q. G., Cai, R. G., Zhang, Y. Z. 2012, PRD, 86, 065004
\bibitem[Guo \& Hu(2013)]{Guo13} Guo, Z. K., Hu, J. W. 2013, PRD, 87, 123519
\bibitem[Hicken et al.(2009)]{Hicken09} Hicken, M., Wood-Vasey, W. M., Blondin, S., et al. 2009, ApJ, 700, 1097
\bibitem[Holsclaw et al.(2010)] {Holsclaw10} Holsclaw, T., Alam, U., Sanso, B., et al. 2010, PRL, 105, 241302
\bibitem[Horava(2009a)]{Horava09a} Horava, P. 2009, JHEP, 03, 020
\bibitem[Horava(2009b)]{Horava09b} Horava, P. 2009, PRD, 79, 084008
\bibitem[Hu \& Sugiyama(1996)]{Hu96} Hu, W., Sugiyama, N. 1996, ApJ, 471, 542
\bibitem[Jacob \& Piran(2007)]{Jacob07} Jacob, U., Piran, T. 2007, Nature Physics, 3, 87
\bibitem[Jacob \& Piran(2008)]{Jacob08} Jacob, U., Piran, T. 2008, JCAP, 0801, 031
\bibitem[Kaaret(1999)]{Kaaret99} Kaaret, P. 1999,  A\&A, 345, L32
\bibitem[Kislat \& Krawczynski(2015)]{Kislat15} Kislat, F., Krawczynski, H. 2015, PRD, 92, 045016
\bibitem[Komatsu et al.(2009)]{Komatsu09} Komatsu, E., Dunkley, J., Nolta, M. R., et al. 2009, ApJS, 180, 330
\bibitem[Komatsu et al.(2011)]{Komatsu11} Komatsu, E., Smith, K. M., Dunkley, J., et al. 2011, ApJS, 192, 18
\bibitem[Kostelecky \& Mewes(2008)]{Kostelecky08} Kostelecky, V. A., Mewes, M. 2008, ApJL, 689, L1
\bibitem[Kostelecky \& Russell(2011)]{Kostelecky11} Kostelecky, V. A., Russell, N. 2011, Rev. Mod. Phys., 83, 11
\bibitem[Kowalski-Glikman \& Nowak(2002)]{Kowalski-Glikman02} Kowalski-Glikman, J., Nowak, S. 2002, PLB, 539, 126
\bibitem[Lewis \& Bridle(2002)]{Lewis02} Lewis, A., Bridle, S. 2002, PRD, 66, 103511
\bibitem[Linder(2003)]{Linder03} Linder, E. V. 2003, PRD, 68, 083504
\bibitem[Mattingly(2005)]{Mattingly05} Mattingly, D. 2005, Living Reviews in Relativity, 8, 5
\bibitem[Magueijo \& Smolin(2002)]{Magueijo02} Magueijo, J., Smolin, L. 2002, PRL, 88, 190403
\bibitem[Pan et al.(2013)]{Pan13} Pan, Y., Cao, S., Gong, Y. G., Liao, K., Zhu, Z. H. 2013, PLB, 718, 699
\bibitem[Pan et al.(2015)]{Pan15} Pan, Y., Gong, Y. G., Cao, S., Gao, H., Zhu, Z. H. 2015, ApJ, 808,1
\bibitem[Passos et al.(2016)]{Passos16} Passos, E., Anacleto, M. A., Brito, F. A., et al. 2016, PLB, 772, 870
\bibitem[Percival et al.(2010)]{Percival10} Percival, B, A., et al. 2010, MNRAS, 401, 2148
\bibitem[Perlmutter et al.(1999)]{Perlmutter99} Perlmutter, S., Aldering, G., Goldhaber, G., et al. 1999, ApJ, 517, 565
\bibitem[Qi et al.(2016)]{Qi16} Qi, J. Z., Zhang, M. J., Liu, W. B., 2016, arXiv:1606.00168
\bibitem[Qi et~al.(2019a)]{Qi19a} Qi, J. Z., Cao, S., Zhang, S. X., et al. 2019, MNRAS, 483, 1104
\bibitem[Ratra \& Peebles(1988)]{Ratra88} Ratra, B., Peebles, P. J. E. 1988, PRD, 37, 3406
\bibitem[Rodriguez Mart{\'{\i}}nez \& Piran (2006)]{Rodriguez06} Rodriguez M, M., Piran, T. 2006, JCAP, 0604, 006
\bibitem[Sarkar(2002)]{Sarkar02} Sarkar, S. 2002, MPLA, 17, 1025
\bibitem[Sefiedgar et al(2011)]{Sefiedgar11} Sefiedgar, A. S., Nozari, K., Sepangi, H. R. 2011, PLB, 696, 119
\bibitem[Seikel et al.(2012)]{SeikeL12a} Seikel, M., Clarkson, C., \& Smith, M. 2012, JCAP, 1206, 036
\bibitem[Spergel et al.(2003)]{Spergel03} Spergel, D. N., Verde, L., Peiris, H. V., et al. 2003, ApJS, 148, 175
\bibitem[Spergel et al.(2007)]{Spergel07} Spergel, D. N., Bean, R., Dor{\'e}, O., et al. 2007, ApJS, 170, 377
\bibitem[Tegmark et al.(2004)]{Tegmark04} Tegmark, M., Blanton, M., Strauss, M., et al. 2004, ApJ, 606, 702
\bibitem[Vasileiou et al.(2015)]{Vasileiou13} Vlasios, V., Jonathan, G., Tsvi, P., Amelino-Camelia, G., 2015, Nature Physics, 11, 344
\bibitem[Vacaru(2010)]{Vacaru10} Vacaru, S. I. 2012, Gen. Rel. Grav, 44, 1015
\bibitem[Wei et al.(2015)]{Wei15} Wei, J. J., Gao, H., Wu, X. F., M{\'e}sz{\'a}ros, P. 2015, PRL, 115, 261101
\bibitem[Wei et al.(2018)]{Wei17a} Zou, X. B., Deng, H. K., Yin, Z. Y., Wei, H. 2018, PLB, 776, 284
\bibitem[Wei et al.(2017a)]{Wei17b} Kyle, F. K., Harriet, L. D., Heeyoung, O., et al. 2017, ApJ, 838, 2
\bibitem[Wei et al.(2017b)]{Wei17c} Wei, J. J., Zhang, B. B., Shao, L., Wu, X. F., M{\'e}sz{\'a}ros, P. 2017, ApJL, 834, L13
\bibitem[Weinberg et al.(2013)]{Weinberg13} Weinberg, D. H., Mortonson, M. J., Eisenstein, D. J., Riess, C. H. J., Rozo, E. 2013, PhR, 530, 87
\bibitem[Xu et al.(2016)]{Xu16} Xu, D., Malesani, D., Fynbo, J. P. U., et al. 2016, GRB Coordinates Network, 19600
\bibitem[Yu \& Wang(2016)]{Yu16} Yu, H., Wang, Y. F. 2016, ApJ, 828, 85
\bibitem[Yang et al.(2015)]{Yang15} Yang, T., Guo, Z. K., Cai, R. G. 2015, PRD, 91, 123533
\bibitem[Zhang \& Xia(2016)]{Zhang16a} Zhang, M. J., Xia, J. Q. 2016, JCAP, 12, 005
\bibitem[Zhang et al.(2016b)]{Zhang16b} Zhang, B. B., Zhang, B., Castro-Tirado, A. J., et al. 2016, arXiv:1612.03089
\bibitem[Zhang \& Ma(2015)]{ZhangS15} Zhang, S., Ma, B. Q. 2015, ApJ, 61, 108
\bibitem[Zhang et al.(2018)]{Zhang18} Zhang, Y., Liu, X. W., Qi, J. Z., Zhang, H. S. 2018, JCAP, 08, 027
\bibitem[Zheng et al.(2019)]{Zheng19} Zheng, X. G., Qi, J. Z., Cao, S., et al. 2019, EPJC, 79, 637
\bibitem[Zou et al.(2018)]{Zou18} Zou, X., Deng, H., Yin, Z., Wei, H. 2018, PLB, 776, 284













\end{thebibliography}
\end{document}